\documentclass[conference]{IEEEtran}
\IEEEoverridecommandlockouts
\usepackage{cite}
\usepackage{amsmath,amssymb,amsfonts}
\usepackage{algorithmic}
\usepackage{graphicx}
\usepackage{textcomp}
\usepackage{xcolor}
\usepackage{subfigure}
\usepackage{algorithm}
\usepackage{algorithmic}

\makeatletter
\newcommand{\removelatexerror}{\let\@latex@error\@gobble}
\makeatother

\ifodd 0
\newcommand{\rev}[1]{{\color{blue}#1}} 
\newcommand{\com}[1]{\textbf{\color{red} (COMMENT: #1)}} 

\else
\newcommand{\rev}[1]{#1}

\newcommand{\com}[1]{}

\fi

\begin{document}
	
	\title{Digital Twin-Assisted Adaptive Preloading for Short Video Streaming\\
	}
	
	\author{
		\IEEEauthorblockN{Shengbo Liu\IEEEauthorrefmark{1}, Wen~Wu\IEEEauthorrefmark{1}, Shaofeng Li\IEEEauthorrefmark{1}, Tom H. Luan\IEEEauthorrefmark{2}, and Xuemin (Sherman) Shen\IEEEauthorrefmark{3}} 
		\IEEEauthorblockA{\IEEEauthorrefmark{1}\textit{Frontier Research Center, Peng Cheng Laboratory, Shenzhen, China}} 
		\IEEEauthorblockA{\IEEEauthorrefmark{2}\textit{School of Cyber Science and Engineering, Xidian University, Xi’an, China}} 
		\IEEEauthorblockA{\IEEEauthorrefmark{3}\textit{Department of Electrical and Computer Engineering, University of Waterloo, Waterloo, Canada}} 
		Email: \{liushb, wuw02, lishf\}@pcl.ac.cn, tom.luan@xidian.edu.cn, sshen@uwaterloo.ca
	}

	\maketitle
	
	\begin{abstract}
		
		\rev{We propose} a digital twin-assisted adaptive preloading scheme to \rev{enhance} bandwidth \rev{efficiency and} user quality of experience (QoE) in short video streaming. We first analyze the relationship between the achievable throughput and video bitrate and critical factors that affect the preloading decision, including the buffer size and bitrate selection. \rev{We then} construct a digital twin-assisted adaptive preloading framework for short video streaming. By collecting and analyzing user’s historical throughput and \rev{tracking} behavior information, a throughput prediction model and a probabilistic model \rev{can be constructed} to accurately predict future throughput and user behavior, respectively. Using the predicted information and real-time running status data from a short video application, we design a preloading strategy to \rev{enhance} bandwidth \rev{efficiency while guaranteeing} user QoE. Simulation results demonstrate \rev{the effectiveness of our proposed scheme comparing with the} state-of-the-art preloading schemes.

	\end{abstract}
	
	
	\section{Introduction}
	
	Nowadays, video streaming occupies around $70\%$ of total mobile data traffic, which is expected to reach $80\%$ by 2028.\footnote{https://www.ericsson.com/en/reports-and-papers/mobility-report/dataforecasts/mobile-traffic-forecast}
	In recent years, short video platforms, e.g., TikTok, YouTube Shorts, and Instagram Reels, have become phenomenal applications \cite{measurement}. The total number of  short video users worldwide has exceeded one billion in 2022 \cite{zhang2022duasvs}. 
	%
	In short video streaming, users can scroll the screen to skip from the current video to the next one in the video recommendation list, allowing them to easily search for their interested videos. Due to channel condition fluctuations in wireless networks \cite{wu2020dynamic}, it is crucial to preload video content on mobile devices to facilitate smooth playback and avoid rebuffering\cite{8845192}. Considering the user's scrolling behavior in short video streaming, it is imperative to preload both the current video and the following ones in the recommended list, which can effectively prevent rebuffering during video playback. However,  the preloaded video can be unwatched due to users' frequent video swiping, which can lead to significant bandwidth waste due to users' frequent video swiping \cite{zhang2021post}. Recent studies indicate that only $30.92\%$ of videos are watched in their entirety \cite{measurement}. Moreover, nearly $45\%$ of the downloaded video is ultimately discarded \cite{network_aware}. Such amount of bandwidth waste not only escalates the cost of mobile data traffic for users but also incurs operational expenses for short video service providers \cite{zhang2022duasvs}. Consequently, an adaptive preloading scheme is essential for short video streaming, aiming to reduce bandwidth waste while enhancing user quality of experience (QoE).

	Recent works have studied the issue of short video preloading in the literature. Nguyen \textit{et al.} considered the network throughput condition and then dynamically adjusted buffer sizes for the current video and the following ones in the video recommendation list\cite{network_aware}. Zhang \textit{et al.} presented a novel preloading mechanism that dynamically preloads the recommended videos to maximize playback smoothness and minimize bandwidth waste \cite{apl2020}. However, these works do not fully consider the impact of varying bitrates on system performance. Recently,  Zhou \textit{et al.} introduced a probabilistic model of user retention to adaptively control the buffer size and the bitrate for preloaded videos \cite{zhou2022pdas}.  Zhang \textit{et al.} proposed a learning-based approach to capture the characteristics of past network conditions and train adaption models for reducing data usage \cite{zhang2022duasvs}. Nevertheless, these works may not fully exploit user-specific information for buffer control and bitrate adaptation.  
	
	To leverage user-specific information and enhance the performance of short video streaming, we introduce the digital twin (DT) technology \cite{shen2021holistic}. DT is a virtual representation of a physical entity that enables real-time synchronization between the digital model and the actual entity \cite{DT_Van}. By applying DT technology, we can create a virtual representation of a user, capturing his/her preferences, behaviors, and viewing patterns. By analyzing this user-specific historical and real-time information, we can make accurate traffic prediction and buffer control and bitrate selection decision, thereby effectively reducing bandwidth waste while ensuring a high-quality user QoE. Therefore, a DT-based solution can dynamically adjust the preloading strategy based on the user's characteristics and real-time network conditions.

	In this paper, we present a DT-assisted adaptive preloading framework for short video streaming, aiming to reduce bandwidth waste and enhance user QoE. We first analyze the relationship between the achievable throughput and the encoding bitrate of a short video. Additionally, we identify the key impact factors on the preloading decision making. Based on these findings, we design a DT framework for short video streaming, consisting of four functional blocks: data repository, throughput prediction, user behavior prediction, and preloading strategy. In particular, the user behavior prediction block constructs a probabilistic model based on the user's historical scrolling behavior information to predict the user's viewing behavior. More importantly, in the preloading strategy block we have proposed effective and low-complexity algorithms to make preloading decisions, including adaptively adjusting buffer sizes for the current video and the ones in the video recommendation list and selecting the bitrates for the upcoming video chunks to download. Extensive simulation results demonstrate the effectiveness of our solution in reducing bandwidth waste and enhancing user QoE, outperforming the state-of-the-art schemes.
	
	
	The main contributions of this paper are summarized as follows:
	\begin{itemize}
		\item We propose a DT framework for short video streaming, which establishes a throughput model and a probabilistic model to predict the user's future throughput and scrolling event.
		\item We design an adaptive short video preloading strategy, which can dynamically adjust the buffer size and adaptively select the bitrate for the current video and the ones in the video recommendation list.
	\end{itemize}

	
	
	The remainder of this paper is organized as follows. Section~\ref{system} introduces the system model. Section \ref{analysis} presents analysis on preloading short videos. The DT-assisted video preloading scheme is presented in Section \ref{DT-framework}. Simulation results are provided in Section \ref{evaluation}, followed by the conclusion in Section \ref{conclusion}.
	
	\section{System Model}\label{system}
	
\begin{figure}[t]
		\centering
		\includegraphics[scale=0.5]{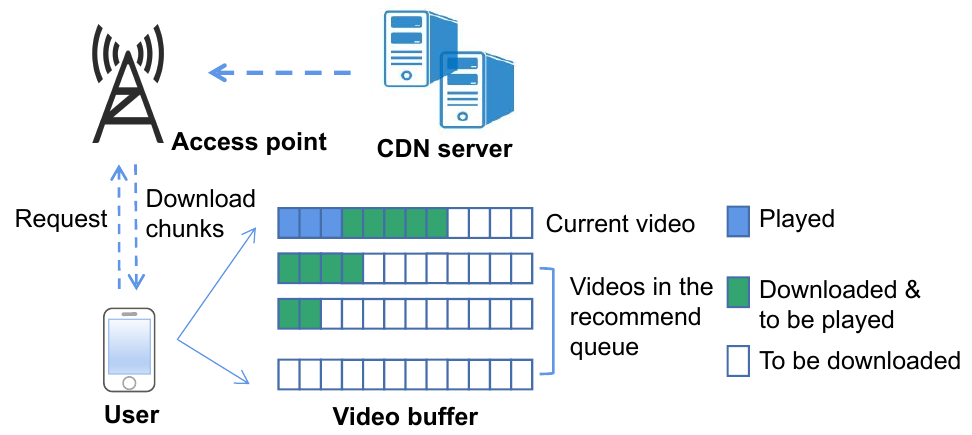}
		\caption{A short video streaming system.}\label{fig_system}
		\vspace{-0.6cm} 
	\end{figure}

	We consider the scenario in which a user utilizes a mobile device to watch videos through a short video application over mobile networks. The general architecture of a short-form video streaming system is depicted in Fig. \ref{fig_system}. In this system, the user initiates video requests directed toward a content delivery network (CDN) server, which serves as a repository for short videos. Each video transmitted to the user's device is segmented into small chunks, featuring consistent playback durations but varying bitrate levels. Upon receipt of video chunks from the CDN server, the user's device stores them in its buffer. To ensure a satisfactory QoE for the user, the CDN server offers a list of recommended videos based on the user's preferences, while the user retains control over which video chunk to preload and determines the appropriate bitrate to accommodate the fluctuating network conditions and the scrolling behavior. It is important to note that the user has the option to preload either the presently viewed video or those featured in the recommended list.

	We consider a viewing session that starts when a user opens the short video application and ends when the user closes the application. During the viewing session, the user watches a list of video ${v_1, v_2, ..., v_N}$ recommended by the CDN server in sequential order. Video $v_i$ is divided by $K_i$ chunks with an identical duration $T_0$. The $k$-th chunk in video $v_i$ is encoded by bitrate $r_{i,k}$, $r_{i,k} \in \mathcal{R}$. 
	
	As discussed in \cite{yin2015control}, user QoE is mainly determined by the average video quality, average quality variations, and the rebuffering time. We define the QoE of the $i$-th video by a weighted sum of the aforementioned factors:
	\begin{align} \label{eq_qoe}
		QoE_i & = \omega_1 \cdot \sum_{k=1}^{K_{i}^s} q(r_{i,k}) - \omega_2 \cdot \sum_{k=1}^{K_{i}^s-1} |q(r_{i,k+1})-q(r_{i,k})| \notag\\ 
		& - \omega_3 \cdot \sum_{k=1}^{K_{i}^s} t_{i,k},
	\end{align}
	where $K_{i}^s$ denotes the number of watched chunks of the $i$-th video, $q(\cdot)$ is the quality metric \cite{yin2015control,zhou2022pdas}, $t_{i,k}$ is the rebuffering time, $\omega_1$, $\omega_2$, and $\omega_3$ are non-negative weighting parameters related to average video quality, video quality variations, and the rebuffering time, respectively. Considering that users may prioritize different factors, the weighting parameters can be customized based on their individual preferences.
	
	The bandwidth consumed by downloading video $i$ is 
	\begin{equation}
		Cost_i = \sum_{k=1}^{K_{i}^d} r_{i,k} \cdot T_0,
	\end{equation}
	where $K_{i}^d$ denotes the number of downloaded chunks of $i$-th video.
	Then, to balance the impacts of bandwidth waste and user QoE on the system performance, we define a utility function U, which is given by
	\begin{equation} \label{eq_goal}
		U = \sum_{i}^{N} (QoE_i - \omega_4 \cdot Cost_i),
	\end{equation}
	where $\omega_4$ denotes the weighting parameter related to the bandwidth usage. Here, the bandwidth wasted at video $i$ due to swiping to the next video is 
	\begin{equation}
		{W_i =  \sum_{k=K_{i}^s+1}^{K_{i}^d} r_{i,k}} \cdot T_0.
	\end{equation}
	
	We can adjust the buffer sizes for the current video and the following ones in the video recommendation list and the bitrate of the upcoming video chunks to be downloaded to maximize the utility function.
	
	\section{Analysis on Short Video Streaming} \label{analysis}
	In this section, we analyze the condition that ensures smooth playback without rebuffering, and the key factors that affect the preloading strategy design.
	
	\subsection{Network Scenario Analysis}
	We consider a worst-case scenario where a user continuously scrolls the screen to swipe the videos once watching the first chunk.  In this scenario, we assume that video playback begins only when the number of buffered chunks reaches an initial threshold denoted as $B_0$, where $B_0 \geq 1$.  To prevent rebuffering events while watching the current video or transitioning to the next video, it is necessary to download the subsequent chunk of the current video and $B_0$ chunks of the next video while the user is viewing the first chunk. Let $C_{min}$ represent the minimum achievable throughput that can sufficiently handle this worst-case scenario. Consequently, $C_{min}$ must satisfy the following equation
	\begin{equation}
		C_{min} = r_{i,1} + \sum_{j=1}^{B_0} r_{i+1,j}.
	\end{equation}
	
	In a simplified scenario where $B_0 = 1$ and all videos are encoded at the same bitrate $R_1$, the minimum throughput threshold is $C_{min} = 2R_1$. This means that the achievable throughput must exceed twice the video bitrate to ensure seamless video playback without rebuffering. In this case, the minimum requirement for video preloading is two chunks.
	
	When the achievable throughput falls below the minimum bitrate $R_{min}$, a rebuffering event will occur. This implies that the download rate is slower than the video's playback rate, resulting in the inability to ensure the smooth playback of the current video. However, if the achievable throughput $C$ satisfies $R_{min} < C < r_{i,1} + \sum_{j=1}^{B_0} r_{i+1,j}$,
	there is room for designing the preloading strategy to prevent rebuffering events.
	
	Based on the aforementioned analysis, we have identified the significance of the achievable throughput in the design of the preloading strategy. Consequently, the following insights can be gleaned:
	\begin{itemize}
		\item When $C \geq \sum_{j=1}^{B_0} r_{i+1,j}$, we can set the minimum preloading threshold to minimize the bandwidth waste;
		\item When $C \leq R_{min}$, there is no necessity to set preloading thresholds and attempt to download the next chunk of the current video;
		\item When $R_{min} < C < \sum_{j=1}^{B_0} r_{i+1,j}$, a preloading strategy must be carefully devised to attain the optimal performance in short-form video streaming.
	\end{itemize}
	
	\subsection{Key Factors}
	Identifying key factors in the design of the preloading strategy is of great significance. Among these factors, the achievable throughput and video bitrate emerge as pivotal  determinants. Additionally, user viewing behavior information such as scrolling pattern, preferred video categories, and minimum viewing time, is crucial in shaping a user's preloading strategy design. Firstly, the scrolling pattern exhibited by a user when using a short-form video application is neither random nor fixed. The preloading strategy of a user who likes frequent swiping videos differs from those who seldom swipe videos. 
	Secondly, users are willing to spend more time watching videos that align with their interests. A study in \cite{zhou2022pdas} shows that approximately $50\%$ of users tend to swipe a category of videos within the first $5\%$ of playback time, while nearly $80\%$ of users complete playback of another category of videos.
	Thirdly, the minimum viewing time of a user is crucial for adjusting the buffer size for current and subsequent videos. By incorporating these factors into preloading decisions, a substantial enhancement in the effectiveness of the preloading strategy can be achieved. 
	

	Considering that the network condition and viewing behavior differ across different users, we adopt DT to facilitate the short video preloading decision by fully extracting useful information from a user's historical and real-time data. Specifically,
	by utilizing a user's historical throughput traces and real-time throughput data, the DT can comprehensively analyze these data and provide accurate throughput prediction. Furthermore, by leveraging a user's historical data on scrolling behavior, combined with information such as video type, viewing progress, and minimum watching duration, the DT can predict the probability of a user's scrolling behavior. With the predicted information, the DT can make high-quality preloading decisions on adjusting the buffer sizes and selecting the bitrates for the upcoming videos to be downloaded. 

	\section{Digital Twin-Assisted Adaptive Preloading Framework} \label{DT-framework}
	
	\begin{figure}[t]
		\centering
		\includegraphics[scale=0.5]{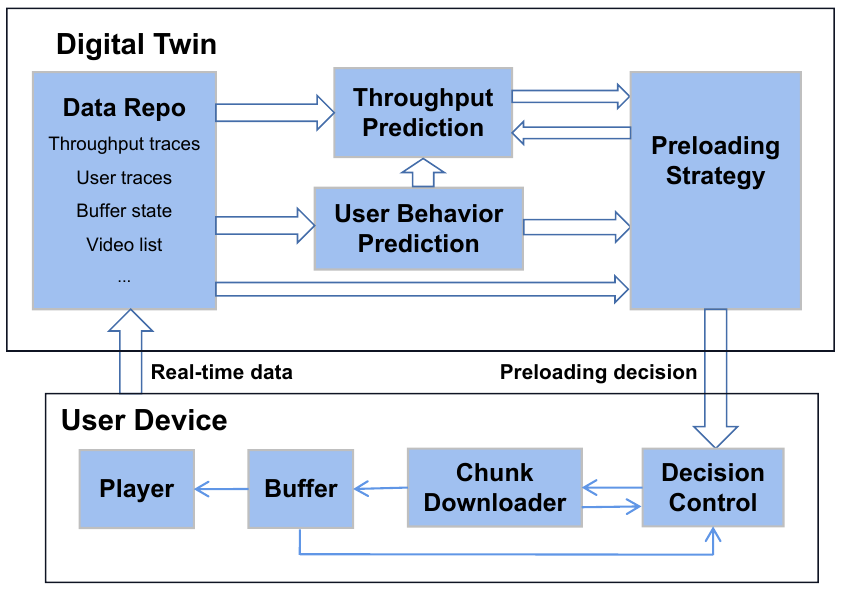}
		\caption{The proposed DTAAP framework.}\label{fig_framework}
		\vspace{-0.6cm} 
	\end{figure}
	In this section, we present a novel framework called Digital Twin-Assisted Adaptive Preloading (DTAAP), aiming to reduce bandwidth waste while ensuring high QoE for users. As depicted in Fig. \ref{fig_framework}, the DTAAP framework leverages the construction of the user's DT to generate a personalized preloading decision by utilizing private user data.
	The DTAAP framework comprises four key functional blocks, namely, data repository, throughput prediction, user behavior prediction, and preloading strategy. Next, we will introduce the four important functional blocks in detail.
	\subsection{Data Repository}
	
	The Data Repository block periodically collects and stores user data related to short video streaming, including throughput traces, user behavior traces, real-time buffer state, video recommendation list, etc. The primary responsibility of this block is to provide the necessary data to the other three blocks. Firstly, the data repository block supplies the throughput prediction block with throughput traces, which is used to predict user future throughput for making informed decisions. Secondly, the data repository block provides the user behavior prediction block with user behavior traces, which is used to analyze user behavior patterns and predict future user actions, facilitating the design of effective preloading strategies. Lastly, the data repository block supplies the buffer state, recommended video list, and other real-time data to the preloading strategy block. This information is essential for determining which video chunk and the corresponding bitrate to preload. 
	
	\subsection{Throughput Prediction}
	Accurate throughput prediction is critical in making effective video preloading decisions, which is expected to prevent rebuffering events caused by network fluctuations. Short video applications are often used in diverse wireless access scenarios, such as open-air spaces, indoor locations, or while on a moving bus. Each scenario presents distinct throughput characteristics, as discussed in \cite{qiao2020trace}. Moreover, the dynamics of the wireless channel and network traffic load contribute to rapid fluctuations in a user's achievable throughput.
	
	Nevertheless, by utilizing the throughput traces provided by the user, the user's digital twin can effectively identify the network environment using the technology proposed in \cite{qiao2020trace}. Once the network scenario is determined, the next step is to design a model that can predict future throughput accurately. One challenge in predicting throughput is the rapid variation experienced during the duration of playing a video chunk, making it difficult to capture this characteristic accurately. To address this issue, we focus on predicting the average throughput over a chunk duration. Additionally, we can leverage bitrate adaptation techniques to mitigate the impact of fast throughput fluctuations. However, it is important to note that frequent bitrate changes can negatively affect user QoE as described by equation \eqref{eq_qoe}. Combining the buffered video chunks and the average throughput of a future period can avoid frequent bitrate changes. Therefore, we use the following model to predict future throughput $C$
	
	\begin{equation}
		C = \alpha_1 \cdot C_{ave}  + \alpha_2 \cdot C_{last},
	\end{equation}
	where $C_{past}$ denotes the past average throughput, $C_{ave}$ denotes the average throughput of downloading last chunk, $\alpha_1$ and $\alpha_2$ are two weighting parameters.

	

	\subsection{User Behavior Prediction}
\begin{figure}[t]
		\centering
		\includegraphics[scale=0.6]{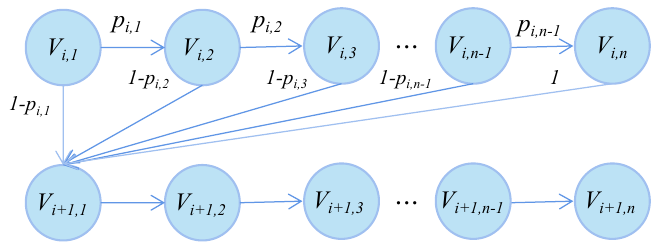}
		\caption{State transition model.}\label{fig_transfer}
		\vspace{-0.6cm} 
	\end{figure}

		\begin{algorithm}[t]\small
		\caption{State transition model construction algorithm. }\label{alg_srd}
		\begin{algorithmic}[1]
			\REQUIRE User traces ${Tr_x(k_x, C_x)}$.
			\ENSURE The probability distribution vector $S$. 
			\STATE Initialize a vector $\textbf{S}=[s_1,s_2,...,s_{100}]$;
			\FOR{$x$ in $1, X$}
			\IF{$\lceil \frac{k_x-1}{K_x} \rceil \neq \lceil \frac{k_x}{K_x} \rceil$}
			\FOR{$y$ in $\lceil \frac{k_x-1}{K_x} \rceil, \lceil \frac{k_x}{K_x} \rceil$}
			\STATE $S[y] = S[y] + 1/X$;
			\ENDFOR
			\ELSE
			\STATE $S[\lceil \frac{k_x}{K_x} \rceil] = S[\lceil \frac{k_x}{K_x} \rceil] + 1/X$;
			\ENDIF
			\ENDFOR
			\RETURN $\textbf{S}$;
		\end{algorithmic}
	\end{algorithm}
	In order to leverage the user's historical scrolling information for better preloading decisions, we construct a state transition model for each video category, as depicted in Fig.~\ref{fig_transfer}. This model allows us to estimate the probabilities associated with watching the next chunk of the same video $p_{i,k}$ and swiping to the next video $1 - p_{i,k}$. Considering that videos within the same category may have different durations, we use $X$ pieces of user trace data $Tr_x(k_x, K_x)$ to represent the scrolling behavior of the $x$-th trace. Here, $k_x$ denotes the viewing chunk number when swiping the video, and $K_x$ represents the total number of chunks in the video. To build state transition model with $1\%$ resolution, we employ Algorithm \ref{alg_srd}. The resulting probability distribution of the model is denoted by a vector $\textbf{S}$. Using this model, we can determine the probability of swiping the video at a specific chunk during playback. For instance, for a video $i$ consisting of $K_i$ chunks, the probability of swiping the video at chunk $v_{i,k}$ can be calculated by
	\begin{equation}
		1-p_{i,k} = \sum_{j = \lceil \frac{k-1}{K_i} \rceil}^{\lceil \frac{k}{K_i} \rceil} S[j].
	\end{equation}
	
	It is worth emphasizing that different categories of videos may have distinct state transition models. By matching the video category with the corresponding model, we can extract valuable information that aids in making informed preloading decisions, including buffer control and bitrate selection. This ensures that the preloading strategy is tailored to the characteristics of each video type, leading to effective and efficient preloading. From the established state transition model,
	we can extract the following valuable insights
	\begin{itemize}
		\item \textbf{Minimum Viewing Duration $K_{min}$}: $K_{min}$ represents the minimum number of video chunks that a user typically watches before deciding whether to swipe the video;
		\item \textbf{Early Scrolling Threshold $K_{early}$}: $K_{early}$ is determined based on the condition $\sum_{j=1}^{K_{early}} (1-p_{i,j}) > P_{th}^{e}$, where $P_{th}^{e}$ is a threshold for early scrolling prediction; 
		\item \textbf{Long Viewing Threshold $K_{long}$}: $K_{long}$ is determined based on the condition $\sum_{j=K_{long}}^{n-1} (1-p_{i,j}) < P_{th}^{l}$, where $P_{th}^{l}$ is a threshold for long viewing prediction.
	\end{itemize}
	

	\subsection{Preloading Strategy}
	The preloading strategy for short video streaming is distinct from that of conventional long video streaming because it involves preloading both the current video and the next videos in the recommended list. The primary goal of the preloading strategy is to minimize bandwidth waste while ensuring a satisfactory QoE for the user. The proposed preloading strategy, outlined in Algorithm \ref{alg_1}, focuses on two essential aspects: buffer control and bitrate adaptation.
	
	\subsubsection{Buffer Control}
	We consider separate buffer thresholds for the current video and the next videos in the recommended list. The number of prefetched video chunks for the current video, denoted as $B_{th}^c$, is determined by balancing the need to reduce rebuffering and reduce bandwidth waste. The value of $B_{th}^c$ is calculated based on the user's viewing behavior and network conditions, and given by

	\begin{small}
	\begin{align} \label{eq_B_c}
	B_{th}^c=\left\{
	\begin{array}{ll}
		1 + \lceil \frac{K_{long}}{K_i} \rceil , & C \geq C_{min} \\
		2 + \lceil \frac{K_{long}}{K_i} \rceil - \lceil \frac{K_{early}}{K_i} \rceil, & R_{i} < C \leq C_{min} \\
		3 + \lceil \frac{K_{long}}{K_i} \rceil - \lceil \frac{K_{early}}{K_i} \rceil, & \text {otherwise }
	\end{array}\right.,
	\end{align}
	\end{small}
    where $R_i$ denotes the bitrate of the downloaded chunk.


	
	For the videos in the recommended list, the number of prefetched video chunks, denoted as $B_{th}^{next}$, is determined to ensure smooth playback and reduce the likelihood of rebuffering when swiping to the next video. The value of $B_{th}^{next}$ is based on factors such as the user's viewing behavior and network conditions, and is given by
	
	\begin{small}
	\begin{align} \label{eq_B_n}
		B_{th}^{next}=\left\{
		\begin{array}{ll}
			1 + K_{min}, & C \geq C_{min} \\
			2 + K_{min} - \lfloor \frac{K_{early}}{2 + K_{min}} \rfloor, & R_{i} < C \leq C_{min} \\
			3 + K_{min} - \lfloor \frac{K_{early}}{3 + K_{min}} \rfloor, & \text { otherwise }
		\end{array}\right..
	\end{align}
	\end{small}
	
	\subsubsection{Bitrate Adaption} 
	To address the challenge of adaptive bitrate selection in response to fluctuating network conditions, we devise an  bitrate adaption scheme that focuses on improving user QoE. The scheme dynamically selects the bitrate of the video chunk to be downloaded, taking into account the available network bandwidth and the buffer state. Algorithm \ref{alg_2} outlines the steps of this scheme. The basic ideas of the scheme can be summarized as follows:
	\begin{itemize}
		\item When a rebuffering occurs, the user lowers the bitrate;
		\item When the buffered video exceeds a threshold, if $C$ is higher than the bitrate of the last downloaded chunk $R_{last}$, the user increases the bitrate to a higher one;
		\item When the buffered video is lower than a threshold, if $C$ is higher than  $R_{last}$, the user keeps preloading video into the buffer; otherwise, the user lowers the bitrate.
	\end{itemize}


	
	\begin{algorithm}[t]\small
		\caption{DT-assisted adaptive preloading algorithm.}\label{alg_1}
		\begin{algorithmic}[1]
			\REQUIRE The current buffer state $B_{i_c}$, the bitrate of last chunk $r_{last}$, the number of preloading videos $N_{pred}$, the sleep time $T_{sleep}$.
			\STATE Update the average throughput $C_{ave}$, the predicted throughput $C_{pred}$, and user behavior;
			\STATE Update $B_{th}^c$ with $C_{pred}$ by \eqref{eq_B_c}; 
			\STATE Update $B_{th}^{next}$ with $C_{ave}$ by \eqref{eq_B_n};
			\IF{$B_{i_c} < B_{th}^c$}
			\STATE Derive $r^*_{i_c,k}$ with Algorithm \ref{alg_2};
			\STATE Download the video data of the current video $v_{i_c}$ with $r^*_{i_c,k}$;
			\ELSE
			\FOR{$j$ in $1, N_{pred}-1$}
			\IF{$B_{i_c+j} < B_{th}^{next}$}
			\STATE Derive $r^*_{i_c+j,k}$ with Algorithm \ref{alg_2};
			\STATE Download the video data for video $v_{i_c+j,k}$ with $r^*_{i_c+j,k}$;
			\ELSIF{$j == N_{pred}-1$}
			\STATE Sleep for $T_{sleep}$ time;
			\ENDIF
			\ENDFOR
			\ENDIF
		\end{algorithmic}
	\end{algorithm}

	\begin{algorithm}[t]\small
		\caption{DT-assisted bitrate adaption algorithm.}\label{alg_2}
		\begin{algorithmic}[1]
			\REQUIRE The average throughput $C_{ave}$, the predicted throughput $C_{pred}$, the buffer state $B_{i_c}$, the bitrate of last chunk $R_{last}$.
			\ENSURE The optimal bitrate $r^*_{i_c+j,k}$.  
			\IF{Preload the current video}
			\IF{A rebuffering event occurs}
			\STATE Reduce the bitrate $r^*_{i_c,k}$ according to $C_{pred}$;
			\ELSIF{$C_{pred} < R_{last}$ and $B_{i_c} < \gamma_1 B_{th}^c$}
			\STATE Reduce the bitrate $r^*_{i_c,k}$ according to $C_{pred}$;
			\ELSIF{$C_{pred} > R_{last}$ and $B_{i_c} > \gamma_2 B_{th}^c$}
			\STATE Increase the bitrate $r^*_{i_c,k}$ according to $C_{pred}$;
			\ENDIF
			\ELSE
			\FOR{$j$ in $1, N_{pred}-1$}
			\STATE Match the bitrate $r^*_{i_c+j,k}$ for the recommended videos $v_{i_c+j}$ with $C_{ave}$;
			\ENDFOR
			\ENDIF
			\RETURN  $r_{i_c,k}^{*}$;
		\end{algorithmic}	
	\end{algorithm}

	

	\section{Performance Evaluation} \label{evaluation}
	\subsection{Experimental Settings}
	We use the trace-driven simulator provided in \cite{zuo2022bandwidth} to evaluate the proposed DTAAP, which simulates the user’s playing and scrolling behavior by sampling the offline video retention rate table and the throughput traces. In this simulator, there are $4$ players in the recommendation list, i.e., $5$ players in total. Each video is encoded into $3$ representations on different bitrates (i.e., $750$kbps, $1200$kbps, and $1850$kbps) and further cropped into chunks with $T_0=1$ second. The test cases are divided into $4$ network scenarios: 1) High bandwidth, 2) medium bandwidth, 3) low bandwidth, and 4) mixed bandwidth. For each case, we sample 50 playback traces, multiplied by 20 network traces to evaluate the performance. Additionally, the weighting parameters for user QoE and the bandwidth usage $\omega_1$, $\omega_2$, $\omega_3$, and $\omega_4$ are respectively set to $1$, $1$, $1.85$, and $0.5$. 
	
	
	\textbf{Compared Bechmarks}. The proposed DTAAP is then compared with the following reference schemes: 
	
	\begin{itemize}
		\item \textbf{Fix-B}: This scheme has a fixed buffer size for the current video and the next videos in the recommended list;
		\item \textbf{NextOne}: This scheme preloads the next video until the current video is fully downloaded \cite{network_aware};
		\item \textbf{Network-based}: This scheme dynamically adjusts the buffer sizes for the current video and the next recommended videos based on the predicted throughput \cite{network_aware};
		\item \textbf{PDAS}: This scheme leverages the user retention probability to build a probabilistic model, which is then utilized to control the maximum buffer size and the bitrate\cite{zhou2022pdas}.
	\end{itemize}
	
	\subsection{Simulation Results}
	
	\begin{figure}[t]
		\graphicspath{{figure/}}
		\centering
		\subfigure[QoE]{
			\includegraphics[width=7cm]{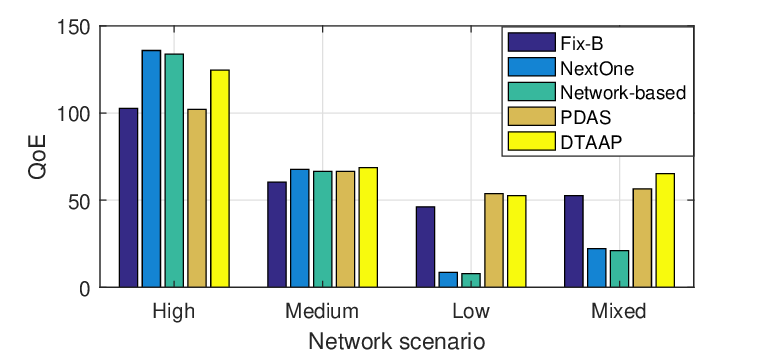}
		}
		\subfigure[Bandwidth waste]{
			\includegraphics[width=7cm]{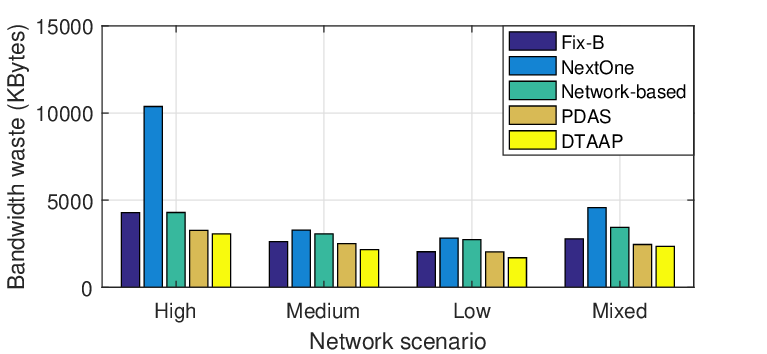}
		}
	\vspace{-0.2cm} 
		\caption{Performance of the proposed scheme and reference schemes.} \label{fig_waste}	
	\vspace{-0.3cm} 
	\end{figure}
	
	\begin{table}[t]
		\centering\caption{Utility Results Achieved on Three Network Scenarios} \vspace{-0.1cm} 
		\label{table_U}
		\begin{tabular}{  l |  c  c  c  c  }
			\hline
			\textbf{Schemes}   & \textbf{High}    & \textbf{Medium}   & \textbf{Low}  & \textbf{Mixed}\\\hline
			Fix-B     & $26.79$  &  $8.24$  & $3.13$ & $-1.92$ \\\hline
			NextOne         & $22.71$ & $-16.78  $    &$ -74.25$ & $-67.91$\\\hline
			Network-based     & $43.87$ & $ -17.35$  &$ -74.79$  & $-64.65$\\\hline
			PDAS     & $30.62$ &  $15.69$ & $11.46$  & $3.34$ \\ \hline
			DTAAP & $44.78$  &  $16.62 $     & $11.41$ & $7.29$  \\\hline
		\end{tabular}
		\vspace{-0.4cm} 
	\end{table}
	
	
	Figure \ref{fig_waste} (a) depicts the QoE performance achieved by the five schemes. It is evident that our proposed DTAAP consistently achieves high QoE values across all four network scenarios. Notably, in the medium and mixed bandwidth scenarios, our DTAAP outperforms the other schemes, attaining the highest QoE levels. Furthermore, in the low bandwidth scenario, our DTAAP achieves the second-best performance, closely approaching the best-performing scheme. These findings underscore the effectiveness of our DTAAP in providing superior user experiences, particularly in challenging network conditions. Additionally, it is observed that user QoE gradually degrades as the available bandwidth decreases.

	
	
	Figure \ref{fig_waste} (b) presents the bandwidth waste incurred by the five schemes evaluated. It is evident that our proposed DTAAP consistently achieves the lowest bandwidth waste across all four network scenarios. The figure also highlights the trend of increasing bandwidth waste as the available bandwidth decreases. This is attributed to the selection of higher bitrates in favorable network conditions, which results in greater data consumption and waste. In comparison to the results depicted in Figure \ref{fig_waste} (a), it is noteworthy that although the NextOne and Network-based schemes achieve higher QoE performance than our DTAAP, they also exhibit higher bandwidth waste. In particular, the NextOne scheme shows significant bandwidth waste due to their full preloading of the current video. This highlights the trade-off between QoE performance and bandwidth efficiency. While these schemes may provide better user experiences, they come at the cost of increased bandwidth waste. In contrast, our DTAAP strikes a balance between QoE performance and bandwidth efficiency. Additionally, the utility results of our DTAAP and the reference schemes are shown in Table \ref{table_U}. It can be observed that our DTAAP consistently achieves optimal performance across all network scenarios.
	
	
	
	

	\section{Conclusion}\label{conclusion} 
	In this paper, we have investigated the adaptive preloading problem in short video streaming with the objective of enhancing user QoE and reducing bandwidth waste. We have proposed a novel digital twin-assisted adaptive preloading framework, named DTAAP, which can leverage the digital twin to capture user network dynamics and scrolling behavior, for enabling adaptive preloading strategy. Extensive simulations demonstrate the effectiveness of DTAAP in enhancing user QoE and reducing bandwidth waste. The DTAAP framework can effectively and efficiently utilize user-speicifc information to enhance the performance of short video streaming in dynamic network environments. For the future work, we will study accurate throughput prediction and effective contextual information utilization for further enhancing user QoE.
	
	

	\bibliographystyle{IEEEtran}
	\bibliography{HNFD}

\end{document}